\documentclass[11pt]{article}

\usepackage[margin=1in,includefoot,footskip=50pt]{geometry}
\usepackage{graphicx}
\usepackage{float}
\usepackage{placeins}
\usepackage{booktabs}
\usepackage{amsmath,amssymb}
\usepackage{array}
\usepackage{microtype}
\usepackage{setspace}
\usepackage{natbib}
\usepackage[hidelinks]{hyperref}
\usepackage{url}
\usepackage{xcolor}

\usepackage{fancyhdr}
\fancypagestyle{firstpage}{%
  \fancyhf{}

  \fancyfoot[L]{%
    \begin{minipage}[t]{0.90\textwidth}
    \setstretch{1}\footnotesize
    \textsuperscript{*}Corresponding author: Mohammad Alamgir Chowdhury\\
    Department of Statistics, University of Manitoba, Winnipeg, Manitoba, Canada\\
    Email: \href{mailto:chowdh62@myumanitoba.ca}{chowdh62@myumanitoba.ca}
    \end{minipage}%
  }
  \fancyfoot[R]{\footnotesize\thepage}
}
\hypersetup{
  pdftitle={Predicting Long-Term ADAS-Cog Change From Early Response After Repetitive Transcranial Magnetic Stimulation in Alzheimer Disease},
  pdfauthor={Mohammad Alamgir Chowdhury and Hina Shaheen}
}

\title{Predicting Long-Term ADAS-Cog Change From Early Response After Repetitive Transcranial Magnetic Stimulation in Alzheimer Disease}

\author{
Mohammad Alamgir Chowdhury$^{a,*}$,
Hina Shaheen$^{b}$\\[0.5em]
\small $^{a}$ Department of Statistics,
University of Manitoba, Winnipeg, Canada\\
\small $^{b}$Faculty of Science, Department of Statistics, 
University of Manitoba, Winnipeg, Canada
}
\date{}
\begin{document}
\maketitle
\thispagestyle{firstpage}

\begin{abstract}
Early cognitive response after repetitive transcranial magnetic stimulation (rTMS) may provide prognostic information about later cognitive change in Alzheimer disease (AD), but it is unclear whether detailed domain-specific response patterns improve prediction beyond simpler changes in the overall Alzheimer Disease Assessment Scale--Cognitive Subscale (ADAS-Cog) total score. This secondary analysis used de-identified data from a randomized rTMS trial comprising 112 participants. A Week-5 landmark cohort included 101 participants, and 91 had complete predictor information and outcomes at 8, 16, and 24 weeks post-treatment. A nested predictor-set comparison evaluated baseline information, trial variables, early total-score response, and early domain-specific response using ridge regression. Ridge regression, multi-task elastic net, restricted random forest, and a longitudinal mixed-effects model were also compared using repeated treatment-site-stratified 5-fold cross-validation with five repetitions and nested hyperparameter tuning. Participant-level paired bootstrap comparisons used 10,000 resamples. Adding Week-3 and Week-5 total-score response substantially improved prediction: the early total-response ridge achieved an overall mean absolute error (MAE) of 3.067, root mean squared error (RMSE) of 3.885, and $R^2=0.384$. Replacing the two total-score response variables with 22 component-level changes worsened performance (MAE 3.363, RMSE 4.269, $R^2=0.257$). The overall paired MAE difference between the total-response and domain-response ridge models was $-0.297$ points (95\% CI $-0.485$ to $-0.115$). The mixed-effects model was competitive (overall MAE 3.162), and its overall difference from the total-response ridge remained uncertain. Sensitivity analyses using absolute future ADAS-Cog scores and excluding Week-3 response supported the main predictive findings. In this sample, early aggregate ADAS-Cog response was more informative for later forecasting than a higher-dimensional domain-specific representation. The model should be interpreted as an internally validated early-response updating tool rather than a pretreatment treatment-selection model or evidence of causal rTMS efficacy.
\end{abstract}


\section{Introduction}

Cognitive decline is a major clinical manifestation of Alzheimer disease (AD)
\citep{scheltens2021alzheimers}. Nonpharmacologic approaches have been investigated
as adjuncts to pharmacologic management of chronic cognitive disorders
\citep{olazaran2010nonpharmacological}. Among these approaches, repetitive
transcranial magnetic stimulation (rTMS) has been studied in mild-to-moderate AD
\citep{Xue2024}. The parent randomized, double-blind, sham-controlled trial included
three intervention schedules: 10 active sessions over 2 weeks, 20 active sessions over
4 weeks, and 4 weeks of sham treatment; ADAS-Cog change was the primary cognitive
outcome \citep{moussavi2021protocol,moussavi2024trial}. The Alzheimer Disease Assessment Scale--Cognitive Subscale (ADAS-Cog) total score
summarizes performance across several cognitive components, but these components
differ in their measurement properties, sensitivity, and responsiveness to change
\citep{cano2010parts,kueper2018adascog,verma2015scoring}. Moreover, the reliability
of ADAS-Cog change scores may be limited, particularly when cognitive change is
evaluated longitudinally \citep{grochowalski2016reliability}. Consequently,
participants with similar changes in the overall ADAS-Cog score may exhibit different
patterns of change across memory, language, orientation, praxis, and other cognitive
domains. This provides a rationale for examining domain-level information when
predicting subsequent cognitive outcomes.

However, the potential prognostic value of domain-specific information should be
distinguished from the predictive value of early cognitive response itself. A simpler
summary of early response is the change in the overall ADAS-Cog total score. If
Week-3 and Week-5 total-score changes contain most of the prognostic information
needed to predict later outcomes, decomposing those changes into individual cognitive
domains may provide little additional predictive benefit while substantially increasing
model complexity. Conversely, if participants with similar total-score responses have
meaningfully different subsequent trajectories depending on which domains changed,
domain-specific measurements could provide incremental prognostic information.
Accordingly, a direct comparison between total-score and domain-specific early-response
models is necessary to determine whether the additional domain-level detail improves
prediction.

Computational approaches have also been used to investigate brain-network dynamics
and disease processes in Alzheimer disease \citep{shaheen_brain_2025}. The present
study addresses a distinct clinical prediction task: forecasting later cognitive outcomes
from measurements already observed during the early treatment period. The objective
is therefore prognostic rather than mechanistic or causal. A decline of approximately three ADAS-Cog points has been proposed as clinically
relevant in patients with early AD \citep{schrag2012clinically}. This emphasizes the
importance of characterizing individual cognitive trajectories, although a threshold
used to define clinically meaningful change is conceptually different from a measure
of prediction error and should not be interpreted as a formal cutoff for acceptable
model performance. Predicting individual cognitive trajectories remains challenging.
Prior rTMS work used baseline electrovestibulography to classify treatment response in
a small cohort \citep{lithgow2022baseline}. Longitudinal ADAS-Cog studies have used
mixed-effects, regularized, tree-based, and multimodal prediction approaches
\citep{Khan2013,prakash2021quantitative}. The TADPOLE challenge found ADAS-Cog13
substantially more difficult to forecast than diagnosis or ventricular volume
\citep{marinescu2021tadpole}. Using data from 1,909 participants with mild cognitive
impairment or AD, Fisher and colleagues modelled 18-month trajectories for 44 clinical
variables \citep{fisher2019forecasting}. Together, these studies demonstrate both the
feasibility and difficulty of longitudinal cognitive prediction. However, they did not
directly examine whether short-term cognitive responses observed after active or sham
rTMS improve later prediction, or whether domain-specific response patterns add
predictive value beyond simpler early changes in the ADAS-Cog total score.

A landmark prediction design establishes a defined time point at which information
observed up to that visit is used to predict outcomes occurring subsequently
\citep{vanhouwelingen2007landmark}. Week 5 provides a clinically interpretable
landmark in the present trial because it follows completion of the planned 2-week or
4-week intervention schedules and allows prediction to incorporate observed early
cognitive response \citep{moussavi2021protocol,moussavi2024trial}. This differs from
pretreatment prediction because each participant's prognosis is updated using
information collected during the early treatment period. Missing outcome data are also a common challenge in longitudinal clinical studies,
particularly when participants miss assessments or discontinue follow-up. As the
number of observed outcomes decreases, estimates may become less precise and may be
biased when the probability of missingness is related to participant characteristics or
clinical outcomes \citep{sterne2009missing,nationalresearchcouncil2010missing}.
Prediction based on information already available by Week 5 therefore represents a
practical longitudinal forecasting problem. Nevertheless, such predictions do not
replace scheduled clinical assessments and cannot eliminate bias arising from
informative missingness.

The present study therefore investigated whether early cognitive response improved
prediction of later ADAS-Cog change and, critically, whether domain-specific response
patterns provided incremental predictive information beyond simpler early total-score
changes. Using Week 5 as the prediction landmark, later ADAS-Cog total-score change
at Weeks 8, 16, and 24 was evaluated through a hierarchy of predictor sets comprising
baseline information, trial-related variables, early Week-3 and Week-5 total-score
response, and additional domain-specific information. This nested comparison directly
tested whether the increased detail of domain-level measurements improved prediction
beyond the overall early cognitive response.

In parallel, four statistical and machine-learning approaches were evaluated for
multi-horizon prediction: ridge regression, multi-task elastic net, restricted random
forest, and a longitudinal mixed-effects model. Secondary objectives were to quantify
uncertainty in paired model comparisons, examine out-of-fold prediction and
calibration, assess the robustness of findings to alternative outcome definitions and
early-response specifications, and characterize the stability of coefficients within the
regularized domain-level model. Because all model development and validation were
performed within a single two-site trial dataset, the study is intended as an internally
validated prognostic analysis rather than evidence of causal treatment effects,
transportability to external populations, or readiness for clinical deployment.

\section{Methods}

\subsection{Study data and cohort construction}

This secondary analysis used de-identified data from a randomized rTMS trial in Alzheimer disease
\citep{moussavi2021protocol,moussavi2024trial,nct02908815}.
The supplied dataset contained 757 rows and 20 variables from 112 participants from two
site-coded strata, MB and MQ. Treatment codes were interpreted according to the parent trial:
R2 represented 2-week active rTMS, R4 represented 4-week active rTMS, and S4 represented
4-week sham. The raw dataset contained both blinded and explicitly labelled unblinded assessment records.
Forty-six rows with assessment labels beginning with ``Unblinded'' were excluded, leaving
711 blinded rows. One participant had a duplicated blinded participant--visit record at the
8-week post-treatment assessment. For this duplicate, ADAS-Cog variables were averaged and
participant-level metadata were retained from the internally consistent records, resulting in
710 unique blinded participant--visit records. Additional one-week assessment labels were not
used because they were outside the prespecified landmark/follow-up structure and contained very
limited ADAS-Cog outcome information. The prediction landmark was the Week-5 assessment. The dataset assessment labels used in the
present analysis were baseline, Week 3, Week 5, and 8, 16, and 24 weeks post-treatment.
The Week-3 assessment had different treatment interpretations across groups: participants in R2
had completed their 2-week active-treatment course, whereas participants in R4 and S4 were still
within their 4-week intervention period. This distinction was considered when interpreting
Week-3 response and was examined in a sensitivity analysis excluding Week-3 response information.
Of the 112 randomized participants, 101 had complete predictor information through the Week-5
landmark. Eleven participants were therefore excluded before the landmark analysis because of
incomplete early predictor information. Among the 101 landmark participants, future outcome
data were available for 99 participants at the 8-week post-treatment horizon, 95 at the
16-week horizon, and 94 at the 24-week horizon. Ninety-one participants had complete outcomes
at all three horizons and constituted the primary three-horizon analysis cohort
(R2, $n=31$; R4, $n=30$; S4, $n=30$).

\subsection{ADAS-Cog variables and outcome construction}

The analysis used the 11-component ADAS-Cog total represented in the dataset as
\texttt{ADAS-Cog-Recog Total Score}. The 11 component scores were recall, naming,
commands, constructional praxis, ideational praxis, orientation, word recognition,
language, comprehension, word finding, and remembering test instructions (RTI).
The corresponding 11-component ADAS-Cog total ranges from 0 to 70, with higher scores
indicating greater cognitive impairment.

Let $X_{ij0}$ denote the baseline score for participant $i$ on ADAS-Cog component $j$,
and let $X_{ij3}$ and $X_{ij5}$ denote the corresponding Week-3 and Week-5 scores.
Early component-level change variables were defined as

\begin{equation}
d3_{ij}=X_{ij3}-X_{ij0},
\end{equation}

and

\begin{equation}
d5_{ij}=X_{ij5}-X_{ij0}.
\end{equation}

Thus, negative values represented improvement relative to baseline and positive values
represented worsening.

Analogous total-score change variables were defined as

\begin{equation}
d3^{\mathrm{total}}_i
=
T_{i3}-T_{i0},
\end{equation}

and

\begin{equation}
d5^{\mathrm{total}}_i
=
T_{i5}-T_{i0},
\end{equation}

where $T_{it}$ denotes the 11-component ADAS-Cog total for participant $i$ at time $t$.

The three prediction outcomes were future ADAS-Cog total changes from baseline:

\begin{equation}
Y_{ih}=T_{ih}-T_{i0},
\qquad
h\in\{8,16,24\},
\end{equation}

where the horizon labels refer to weeks post-treatment in the supplied dataset. Under this
definition, negative future change represents cognitive improvement relative to baseline and
positive change represents worsening.

For the domain-rich supervised models, the numeric predictor set consisted of age,
the 11 baseline ADAS-Cog component scores, the 11 Week-3 component changes, and the
11 Week-5 component changes, giving 34 numeric predictors. Sex, treatment group, and
site were included as categorical predictors, resulting in 37 predictors before encoding. Baseline ADAS-Cog total was not included simultaneously with the 11 baseline component scores
in the domain-rich supervised models because the total was deterministically reconstructed
from those component scores. After one-hot encoding, the domain-rich design matrix contained
38 columns. The reference categories were Female for sex, R2 for treatment, and MB for site.

The primary complete-case cohort contained no missing values in the predictor sets used for
model fitting; therefore, no predictor imputation was performed. Numeric variables were
standardized using means and standard deviations estimated exclusively from the relevant
training fold. Categorical variables were one-hot encoded with the first level omitted.
All preprocessing was embedded within the cross-validation pipelines to prevent information
leakage. Spearman rank correlations were calculated in the complete-case cohort to characterize
monotonic dependence among the study variables. Two descriptive correlation matrices were
examined: correlations among the 11 baseline ADAS-Cog component scores and correlations among
the three future ADAS-Cog change outcomes. The correlation analyses were not used for predictor selection, model tuning, or performance
evaluation. Baseline-domain correlations were used to characterize overlap among cognitive
components, whereas correlations among the three future outcomes were used to quantify the
degree to which the prediction horizons contained shared information.

\subsection{Predictor-set comparison}

A ridge-regression predictor-set comparison was conducted to distinguish the predictive value
of baseline information, trial-design variables, aggregate early cognitive response, and
domain-specific early response. All four specifications used the same 91 participants,
outcomes, outer cross-validation splits, preprocessing procedures, and ridge-regression
tuning strategy.

The baseline-only ridge model included

\begin{equation}
\mathcal{M}_{A}
=
\{
\mathrm{age},
\mathrm{sex},
T_{0}
\}.
\end{equation}

The baseline-plus-trial ridge model additionally included treatment group and site:

\begin{equation}
\mathcal{M}_{B}
=
\mathcal{M}_{A}
+
\{
\mathrm{treatment},
\mathrm{site}
\}.
\end{equation}

The early total-response ridge additionally represented early cognitive response using
the Week-3 and Week-5 total-score changes:

\begin{equation}
\mathcal{M}_{C}
=
\mathcal{M}_{B}
+
\{
d3^{\mathrm{total}},
d5^{\mathrm{total}}
\}.
\end{equation}

This specification contained seven predictors before categorical encoding.

The early domain-response ridge used the same baseline and trial covariates as
$\mathcal{M}_{C}$ but replaced the two total-score change variables with the 11 Week-3
and 11 Week-5 component-specific changes:

\begin{equation}
\mathcal{M}_{D}
=
\mathcal{M}_{B}
+
\{d3_j\}_{j=1}^{11}
+
\{d5_j\}_{j=1}^{11}.
\end{equation}

This specification contained 27 predictors before categorical encoding.

The early domain-response model did not additionally contain
$d3^{\mathrm{total}}$ or $d5^{\mathrm{total}}$, because those quantities are constructed
from the same ADAS-Cog components. Baseline component scores were also omitted from
$\mathcal{M}_{D}$ so that the baseline information supplied to the total-response and
domain-response specifications was held constant. Consequently, the comparison between
$\mathcal{M}_{C}$ and $\mathcal{M}_{D}$ evaluated whether representing early response
through individual ADAS-Cog components improved prediction beyond the simpler aggregate
total-score representation.

Models $\mathcal{M}_{A}$--$\mathcal{M}_{C}$ form a hierarchical sequence, whereas
$\mathcal{M}_{C}$ and $\mathcal{M}_{D}$ are matched alternative representations of
early cognitive response rather than strictly nested models.

\subsection{Model-class comparison}

A separate model-class comparison evaluated four approaches using richer early cognitive
information: full-domain multi-output ridge regression, multi-task elastic net,
a restricted multi-output random forest, and a longitudinal mixed-effects model. The full-domain ridge, elastic-net, and random-forest models used age, all 11 baseline
ADAS-Cog components, all 11 Week-3 component changes, all 11 Week-5 component changes,
sex, treatment group, and site. Thus, these three supervised approaches received the same
domain-rich predictor information. The mixed-effects model used a reduced longitudinal representation based on grouped
cognitive composites rather than all 11 components individually. The mixed-effects
comparison was therefore interpreted as a comparison of modelling strategies and predictor
representations rather than as a pure algorithm comparison using an identical design matrix.

\subsection{Ridge regression}

For both the predictor-set analysis and the full-domain ridge model, multi-output ridge
regression was fitted to the three future outcomes simultaneously. Let
$\mathbf{Y}\in\mathbb{R}^{n\times 3}$ denote the matrix of future changes and
$\mathbf{X}$ the predictor matrix. Ridge regression estimated coefficient matrix
$\mathbf{B}$ by minimizing

\begin{equation}
\left\|
\mathbf{Y}-\mathbf{X}\mathbf{B}
\right\|_{F}^{2}
+
\alpha
\left\|
\mathbf{B}
\right\|_{F}^{2},
\end{equation}

with an intercept fitted separately. A common penalty parameter $\alpha$ was applied
across the three outcomes and selected from

\begin{equation}
\alpha
\in
\{0.01,0.1,1,10,100,1000\}.
\end{equation}

Predictors were standardized within training folds, whereas outcome variables were not
standardized. Hyperparameter selection minimized the equally weighted mean absolute error
across the three prediction horizons.

\subsection{Multi-task elastic net}

The multi-task elastic net jointly estimated coefficients for all three outcomes and
encouraged shared predictor selection across horizons \citep{zou2005elasticnet}. Its
objective was

\begin{equation}
\frac{1}{2n}
\left\|
\mathbf{Y}-\mathbf{X}\mathbf{B}
\right\|_{F}^{2}
+
\alpha\rho
\left\|
\mathbf{B}
\right\|_{2,1}
+
\frac{\alpha(1-\rho)}{2}
\left\|
\mathbf{B}
\right\|_{F}^{2},
\end{equation}

where $\rho$ is the L1 ratio and
$\|\mathbf{B}\|_{2,1}$ denotes the sum of the row-wise Euclidean norms of the coefficient
matrix. This group penalty encourages a predictor to enter or leave the three outcome
models jointly.

The candidate values were

\begin{equation}
\alpha
\in
\{0.003,0.01,0.03,0.1,0.3,1\},
\end{equation}

and

\begin{equation}
\rho
\in
\{0.1,0.5,0.9\}.
\end{equation}

An intercept was fitted, the optimization tolerance was $10^{-4}$, the maximum number of
iterations was 50,000, and coordinate updates used cyclic selection. Outcomes were not
standardized.

\subsection{Restricted random forest}

A single multi-output random-forest regressor was fitted to the three outcomes.
Each forest contained 200 trees. Hyperparameter tuning considered

\begin{align}
\mathrm{maximum\ depth}
&\in
\{2,3,4,5,\mathrm{unrestricted}\},\\
\mathrm{minimum\ leaf\ size}
&\in
\{2,3,5,8,10\},\\
\mathrm{minimum\ split\ size}
&\in
\{2,5,10\},\\
\mathrm{maximum\ feature\ fraction}
&\in
\{0.3,0.5,0.8,1.0\}.
\end{align}

Twelve candidate hyperparameter combinations were evaluated by randomized search within each
outer training set. The forest was intentionally restricted because of the relatively small
sample size and the risk of overfitting with highly flexible tree ensembles
\citep{breiman2001random}.

\subsection{Longitudinal mixed-effects model}

A longitudinal mixed-effects model was included as a secondary comparator to account
explicitly for the repeated future outcomes observed within each participant. The three
future ADAS-Cog change outcomes were reshaped into long format, resulting in three
records per participant corresponding to the 8-, 16-, and 24-week post-treatment
prediction horizons. Because fitting all 11 individual domain changes within a longitudinal mixed-effects
structure would have introduced a comparatively large number of parameters relative to
the sample size, early domain changes were summarized into three cognitive composites.
The memory composite was defined as the sum of recall, orientation, word recognition,
and remembering-test-instructions (RTI) changes. The language composite was defined as
the sum of naming, commands, language, comprehension, and word-finding changes. The
praxis composite was defined as the sum of constructional-praxis and ideational-praxis
changes. Separate Week-3 and Week-5 versions of each composite were constructed.

The model contained both fixed and random effects. The fixed effects represented
population-average associations shared across participants, whereas the random effect
represented participant-specific deviation from the population-average intercept. The fixed-effects portion of the model included prediction horizon, treatment group,
the horizon-by-treatment interaction, age, baseline ADAS-Cog total score, Week-3 and
Week-5 memory composites, Week-3 and Week-5 language composites, Week-3 and Week-5
praxis composites, sex, and study site. Continuous predictors were standardized using
means and standard deviations estimated from the corresponding outer training fold. Prediction horizon was entered as a three-level categorical variable corresponding
to 8, 16, and 24 weeks post-treatment, with the 8-week horizon used as the reference
level. Treatment group was entered as a categorical variable, and the
horizon-by-treatment interaction allowed treatment-group differences to vary across
prediction horizons.

The model can be written as

\begin{align}
Y_{ih}
={}&
\beta_0
+
\beta_{\mathrm{horizon},h}
+
\beta_{\mathrm{treatment},i}
+
\beta_{\mathrm{horizon}\times\mathrm{treatment},ih}
\nonumber\\
&+
\beta_{\mathrm{age}}z(\mathrm{age}_i)
+
\beta_{\mathrm{base}}z(T_{i0})
\nonumber\\
&+
\beta_{3M}z(d3_{\mathrm{memory},i})
+
\beta_{5M}z(d5_{\mathrm{memory},i})
\nonumber\\
&+
\beta_{3L}z(d3_{\mathrm{language},i})
+
\beta_{5L}z(d5_{\mathrm{language},i})
\nonumber\\
&+
\beta_{3P}z(d3_{\mathrm{praxis},i})
+
\beta_{5P}z(d5_{\mathrm{praxis},i})
\nonumber\\
&+
\beta_{\mathrm{sex},i}
+
\beta_{\mathrm{site},i}
+
u_i
+
\epsilon_{ih},
\end{align}

where $Y_{ih}$ denotes future ADAS-Cog change for participant $i$ at horizon $h$,
$T_{i0}$ denotes baseline ADAS-Cog total score, and $z(\cdot)$ denotes
training-fold standardization.

The fixed effects were therefore prediction horizon, treatment group,
horizon-by-treatment interaction, age, baseline ADAS-Cog total score,
Week-3 and Week-5 memory change, Week-3 and Week-5 language change,
Week-3 and Week-5 praxis change, sex, and site.

The only random effect was a participant-specific random intercept,

\begin{equation}
u_i \sim N(0,\sigma_u^2).
\end{equation}

The random intercept allowed each participant to have an individual deviation from
the population-average intercept and accounted for correlation among the three
future observations contributed by the same participant. No participant-specific
random slopes were fitted. A random-intercept-only structure was used for parsimony
because each participant contributed only three future outcomes and the model was
used primarily as an internally validated prediction comparator.

Residual errors were represented as

\begin{equation}
\epsilon_{ih} \sim N(0,\sigma_{\epsilon}^{2}),
\end{equation}

with residuals assumed independent conditional on the participant-specific random
intercept and fixed effects. Within-participant dependence across the three horizons
was therefore induced by the shared random intercept rather than by a separate
horizon-specific residual covariance structure. The model was fitted using maximum likelihood rather than restricted maximum
likelihood (REML) with \texttt{statsmodels.MixedLM}. The L-BFGS optimizer was
attempted first, followed by Powell and conjugate-gradient optimization if needed,
with a maximum of 2,000 iterations. All 25 outer-cross-validation fits converged
using L-BFGS. For held-out participants, predictions were generated from the fixed-effects
component only. Participant-specific random intercepts could not be estimated for
individuals who were absent from the training data because no outcome observations
for those individuals were available at prediction time. Using the held-out outcomes
to estimate participant-specific random effects would have introduced information
leakage. Consequently, the cross-validated mixed-effects predictions represented
population-level predictions for previously unseen participants.

\subsection{Cross-validation and hyperparameter tuning}

Participants were kept intact throughout model development and evaluation. The primary
validation scheme used five-fold cross-validation repeated five times, yielding 25 outer
training/test splits. The random seed was 20260822. Outer folds were jointly stratified by treatment group and site. In the complete-case cohort,
the six treatment-site strata contained 19 R2-MB, 12 R2-MQ, 16 R4-MB, 14 R4-MQ,
19 S4-MB, and 11 S4-MQ participants. Outer training sets therefore contained 72 or 73
participants and test sets contained 18 or 19 participants. The same outer participant splits were used for all supervised models, the predictor-set
comparisons, and the mixed-effects comparator. Hyperparameters for the supervised models
were selected exclusively within each outer training set using three-fold stratified inner
cross-validation. Inner folds used the same treatment-site stratification and a deterministic
seed derived from the primary seed and outer-split number. The inner tuning criterion was negative mean absolute error as implemented in
scikit-learn. For multi-output predictions, MAE was averaged equally across the three
outcome columns; therefore, the 8-, 16-, and 24-week prediction horizons received equal
weight during tuning. Nested tuning was used to reduce selection-induced optimism
\citep{cawley2010overfitting}.

\subsection{Performance metrics}

Prediction performance was evaluated using mean absolute error (MAE), root mean squared error
(RMSE), and coefficient of determination ($R^2$).

For horizon $h$,

\begin{equation}
\mathrm{MAE}_h
=
\frac{1}{n}
\sum_{i=1}^{n}
\left|
y_{ih}-\hat{y}_{ih}
\right|,
\end{equation}

and

\begin{equation}
\mathrm{RMSE}_h
=
\sqrt{
\frac{1}{n}
\sum_{i=1}^{n}
\left(
y_{ih}-\hat{y}_{ih}
\right)^2
}.
\end{equation}

Horizon-specific $R^2$ was calculated as

\begin{equation}
R_h^2
=
1-
\frac{
\sum_{i=1}^{n}
(y_{ih}-\hat{y}_{ih})^2
}{
\sum_{i=1}^{n}
(y_{ih}-\bar{y}_{h})^2
},
\end{equation}

where $\bar{y}_{h}$ denotes the observed mean in the corresponding evaluation data rather
than the training-sample mean.

Overall MAE and RMSE were calculated using equal weighting of the three outcome columns.
Overall $R^2$ used variance-weighted multi-output aggregation, such that horizons with
greater observed outcome variability received proportionally greater weight. Performance was calculated separately for each of the five repeated outer cross-validation
runs and then summarized by the mean and standard deviation. These standard deviations
describe variability across repeated train/test partitions and were not interpreted as
independent standard errors or confidence intervals.

\subsection{Paired bootstrap comparison}

Participant-level paired bootstrap resampling was used to quantify uncertainty in prediction
error differences using the repeated out-of-fold predictions. The primary bootstrap comparison used the early total-response ridge,
$\mathcal{M}_{C}$, as the reference model and compared it with the early domain-response
ridge, $\mathcal{M}_{D}$, and the longitudinal mixed-effects model. These comparisons
addressed whether representing early cognitive response through individual domains improved
prediction over aggregate total-score response and whether the numerically strongest
ridge specification differed meaningfully from the longitudinal mixed-effects comparator.

For each participant and prediction horizon, absolute error was calculated for each of the
five repeated out-of-fold predictions and then averaged across repetitions. For the overall
comparison, participant-level absolute errors were additionally averaged across the three
prediction horizons. Participants were resampled with replacement while retaining all three prediction horizons
for each selected participant. Ten thousand bootstrap samples were generated using random
seed 20260827. Previously obtained out-of-fold predictions were resampled; models were not
refitted within individual bootstrap samples. For the primary comparisons, MAE differences were defined as

\begin{equation}
\Delta_{\mathrm{MAE}}
=
\mathrm{MAE}_{\mathcal{M}_{C}}
-
\mathrm{MAE}_{\mathrm{comparator}},
\end{equation}

so that negative values favored the early total-response ridge and positive values favored
the comparator. Percentile 95\% confidence intervals and the bootstrap probability that the
early total-response ridge had lower MAE were calculated separately for each horizon and
overall.

A secondary bootstrap analysis was also performed for the domain-rich model-class comparison.
Full-domain ridge regression was used as the reference and was compared with multi-task
elastic net, restricted random forest, and the mixed-effects model. In this analysis,
differences were defined as

\begin{equation}
\Delta_{\mathrm{MAE}}^{\mathrm{domain}}
=
\mathrm{MAE}_{\mathrm{full\mbox{-}domain\ ridge}}
-
\mathrm{MAE}_{\mathrm{comparator}},
\end{equation}

with negative values favoring full-domain ridge regression. The same participant-level
resampling procedure and 10,000 bootstrap samples were used.

\subsection{Sensitivity analyses}

Sensitivity analyses were conducted to evaluate assumptions of the primary analysis. Because the early change predictors and the future change outcomes were both constructed
relative to baseline, a sensitivity analysis used absolute follow-up ADAS-Cog total scores
at the three future horizons as the outcomes. The domain-response ridge predictor set,
including baseline ADAS-Cog total, age, sex, treatment, site, and Week-3 and Week-5
component-level changes, was used for this analysis. This alternative outcome formulation
removed baseline subtraction from the outcome and therefore reduced concern that predictive
performance arose primarily from mathematical coupling through a shared baseline term.

A second sensitivity analysis excluded all Week-3 response information because Week 3
occurred after completion of treatment for R2 but during the intervention period for R4
and S4. The Week-3-excluded specification used age, sex, baseline ADAS-Cog total,
treatment group, site, Week-5 total-score change, the 11 baseline ADAS-Cog component scores,
and the 11 Week-5 component changes. This analysis was interpreted as a broader
Week-5-focused sensitivity analysis rather than as a strictly nested deletion from
$\mathcal{M}_{D}$. Characteristics of participants included in the complete three-horizon cohort were also
compared descriptively with those excluded because of incomplete landmark predictor
information or incomplete future outcomes. These comparisons were used to characterize
potential differences associated with the complete-case restriction and were not used
for inferential testing.

\subsection{Calibration and coefficient stability}

Calibration was evaluated for the early total-response ridge model,
$\mathcal{M}_{C}$. Each participant had one out-of-fold prediction at each horizon
within each of the five cross-validation repetitions. For each participant and horizon,
the five out-of-fold predictions were averaged to obtain a single cross-validated
prediction. Calibration was then assessed separately at the 8-, 16-, and 24-week horizons by
regressing the observed ADAS-Cog change on the corresponding mean out-of-fold prediction
using ordinary least squares:

\begin{equation}
Y_{ih}
=
\gamma_{0h}
+
\gamma_{1h}\hat{Y}_{ih}
+
e_{ih}.
\end{equation}

The calibration intercept $\gamma_{0h}$ and slope $\gamma_{1h}$ were reported with
95\% confidence intervals. Ideal calibration corresponds to an intercept of zero and
a slope of one. Observed-versus-predicted plots included both the identity line and
the estimated calibration line. For exploratory interpretation of the full-domain ridge model, ridge regression was
refitted within each outer training set using the penalty parameter selected by the
corresponding inner cross-validation procedure. Numeric coefficients were expressed per
one training-fold standard deviation. For each predictor and horizon, the mean coefficient,
standard deviation, mean absolute coefficient, and sign consistency were summarized across
the 25 outer fits.

Sign consistency was defined as

\begin{equation}
\max
\left[
P(\hat{\beta}>0),
1-P(\hat{\beta}>0)
\right].
\end{equation}

A value of 1 indicated that the coefficient had the same sign in every outer fit, and
a threshold of 0.80 was used descriptively to identify coefficients with relatively stable
direction. For supplementary visualization, the 12 numeric predictors with the largest
average absolute standardized coefficients across the outer fits were displayed.
These coefficient summaries were treated as exploratory conditional predictive associations
and not as causal effects or evidence of independent biological importance.

Analyses were performed in Python using pandas 3.0.3, NumPy 2.4.6,
scikit-learn 1.9.0, statsmodels 0.14.6, SciPy 1.18.0,
Matplotlib 3.11.0, and seaborn 0.13.2
\citep{pedregosa2011scikit,seabold2010statsmodels}.
Random seeds, cross-validation splits, preprocessing operations, model specifications,
hyperparameter grids, and bootstrap procedures were fixed programmatically to support
reproducibility.

\FloatBarrier
\section{Results}

Of the 112 randomized participants, 101 had complete baseline, Week-3,
and Week-5 predictor information and therefore entered the Week-5
landmark cohort. Of these, 91 participants had observed ADAS-Cog outcomes
at all three future horizons and constituted the primary complete-case
prediction cohort. The final cohort contained 31 participants in R2,
30 in R4, and 30 in S4. Fifty-four participants were from site MB and
37 from site MQ.

\begin{figure}[htbp]
\centering
\includegraphics[width=0.90\textwidth,height=0.60\textheight,keepaspectratio]{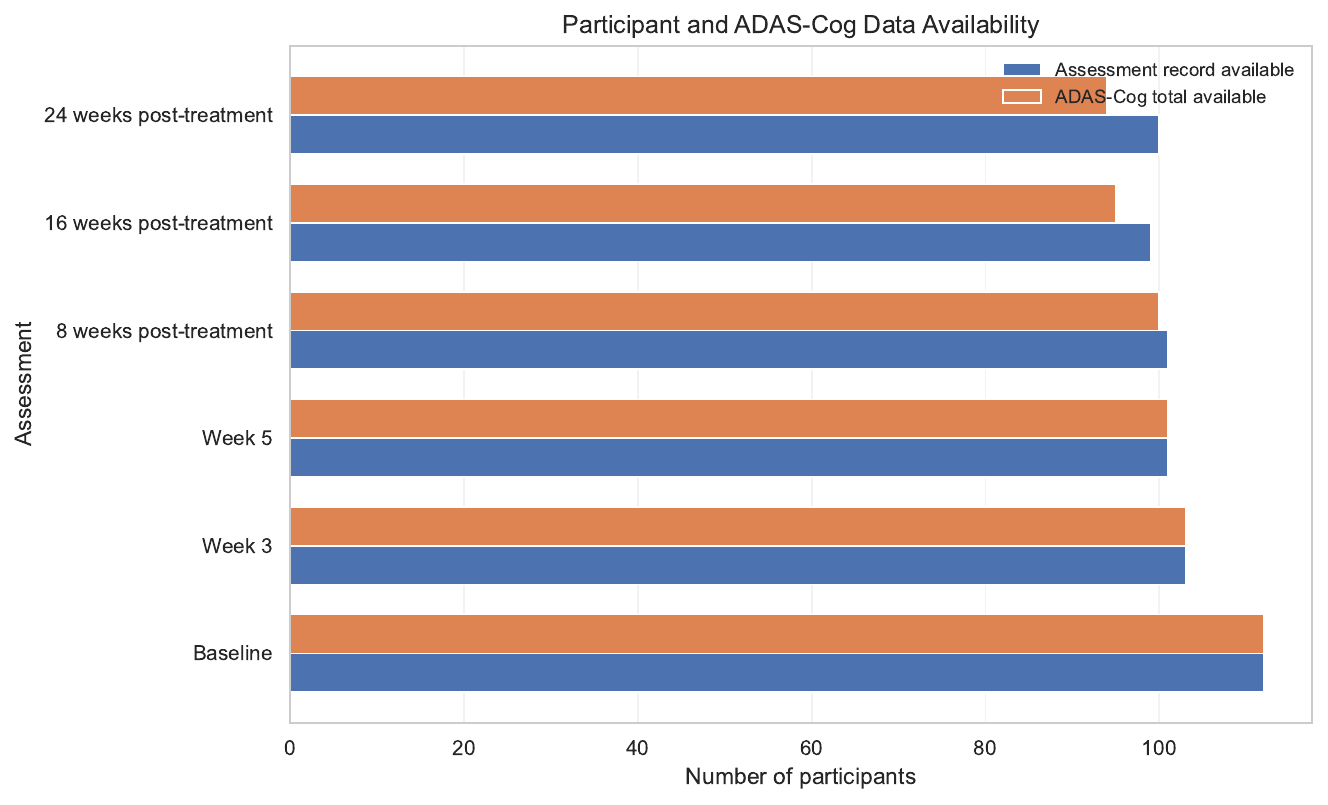}
\caption{Participant and ADAS-Cog data availability across the assessment schedule.}
\label{fig:availability}
\end{figure}

Among the 101 landmark participants, future ADAS-Cog outcomes were
available for 99 participants at 8 weeks post-treatment, 95 at 16 weeks,
and 94 at 24 weeks. Eleven participants were excluded before the landmark
because of incomplete early predictor information, and an additional
10 landmark participants were excluded from the primary three-horizon
analysis because at least one future outcome was missing
(Table~\ref{tab:cohort-flow}; Figure~\ref{fig:availability}).

\begin{table}[t]
\centering
\caption{Participant flow and primary analysis cohort.}
\label{tab:cohort-flow}
\begin{tabular}{lr}
\toprule
Analysis stage & Participants \\
\midrule
Randomized / baseline cohort & 112 \\
Complete Week-5 landmark predictors & 101 \\
Observed 8-week outcome among landmark cohort & 99 \\
Observed 16-week outcome among landmark cohort & 95 \\
Observed 24-week outcome among landmark cohort & 94 \\
Complete all three future outcomes & 91 \\
\midrule
R2 in primary cohort & 31 \\
R4 in primary cohort & 30 \\
S4 in primary cohort & 30 \\
MB in primary cohort & 54 \\
MQ in primary cohort & 37 \\
\bottomrule
\end{tabular}
\end{table}

Participants included in the primary analysis had a mean age of
72.48 years (SD 8.41) and a mean baseline 11-component ADAS-Cog total
score of 23.94 (SD 9.02). The 21 participants excluded from the primary
complete-case analysis had a mean age of 75.71 years (SD 7.04) and a
mean baseline ADAS-Cog total score of 26.65 (SD 8.94). Among included
participants, 38 were female and 53 were male; among excluded participants,
14 were female and seven were male.

Mean future ADAS-Cog change varied across treatment groups and horizons
(Table~\ref{tab:outcome-change}). Negative values represent improvement
relative to baseline and positive values represent worsening. At the
8-week post-treatment horizon, mean change was negative in all three
treatment groups. By 24 weeks, mean change was positive in all three
groups. These summaries are descriptive and were not used to estimate
treatment effects.

\begin{figure}[htbp]
\centering
\includegraphics[width=\textwidth,height=0.60\textheight,keepaspectratio]{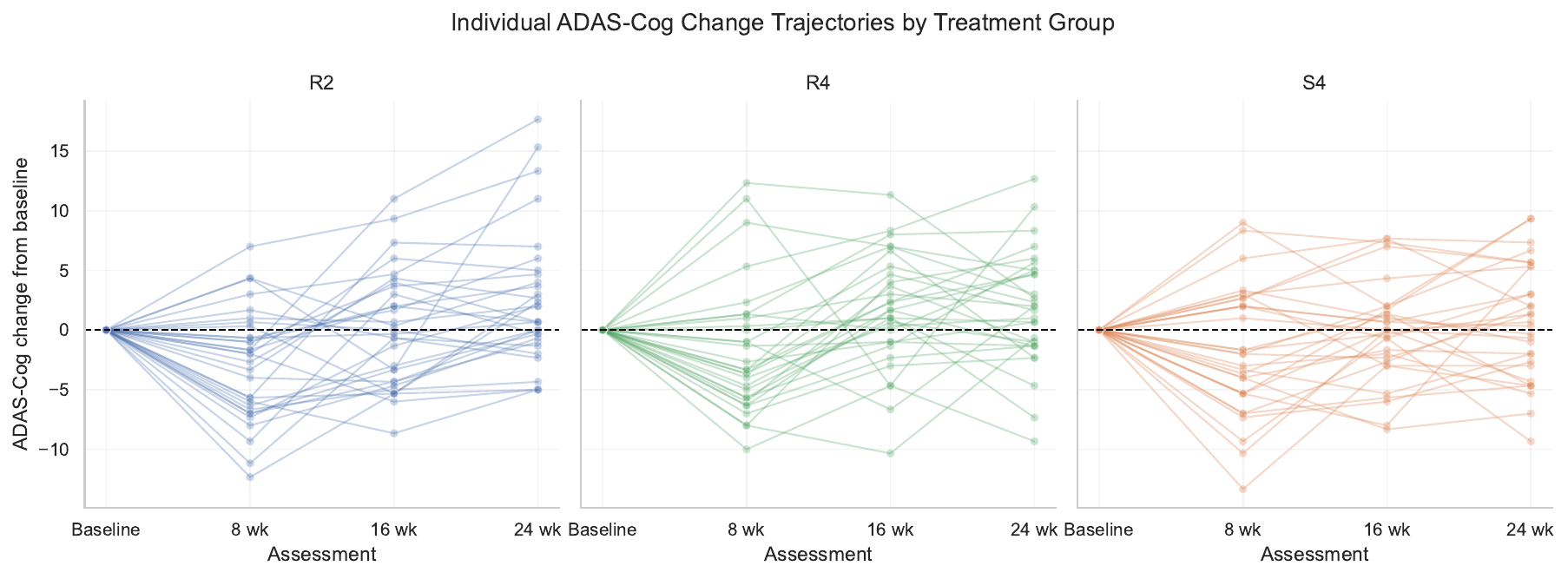}
\caption{Individual ADAS-Cog change trajectories by treatment group.
Negative change indicates improvement relative to baseline and positive
change indicates worsening.}
\label{fig:trajectories}
\end{figure}

\begin{table}[t]
\centering
\caption{ADAS-Cog total-score change from baseline by treatment group
in the primary complete-case cohort. Values are mean (SD).}
\label{tab:outcome-change}
\begin{tabular}{lccc}
\toprule
Treatment & 8 weeks & 16 weeks & 24 weeks \\
\midrule
R2 & $-2.930$ (4.651) & $-0.053$ (4.839) & 2.397 (5.675) \\
R4 & $-1.756$ (5.509) & 1.534 (4.753) & 1.755 (4.809) \\
S4 & $-1.690$ (5.398) & $-0.167$ (4.246) & 0.555 (4.873) \\
\bottomrule
\end{tabular}
\end{table}
Individual outcome trajectories showed substantial within-group
heterogeneity, with both improvement and worsening observed within each
treatment group (Figure~\ref{fig:trajectories}).

\begin{figure}[htbp]
\centering
\includegraphics[width=0.82\textwidth,height=0.60\textheight,keepaspectratio]{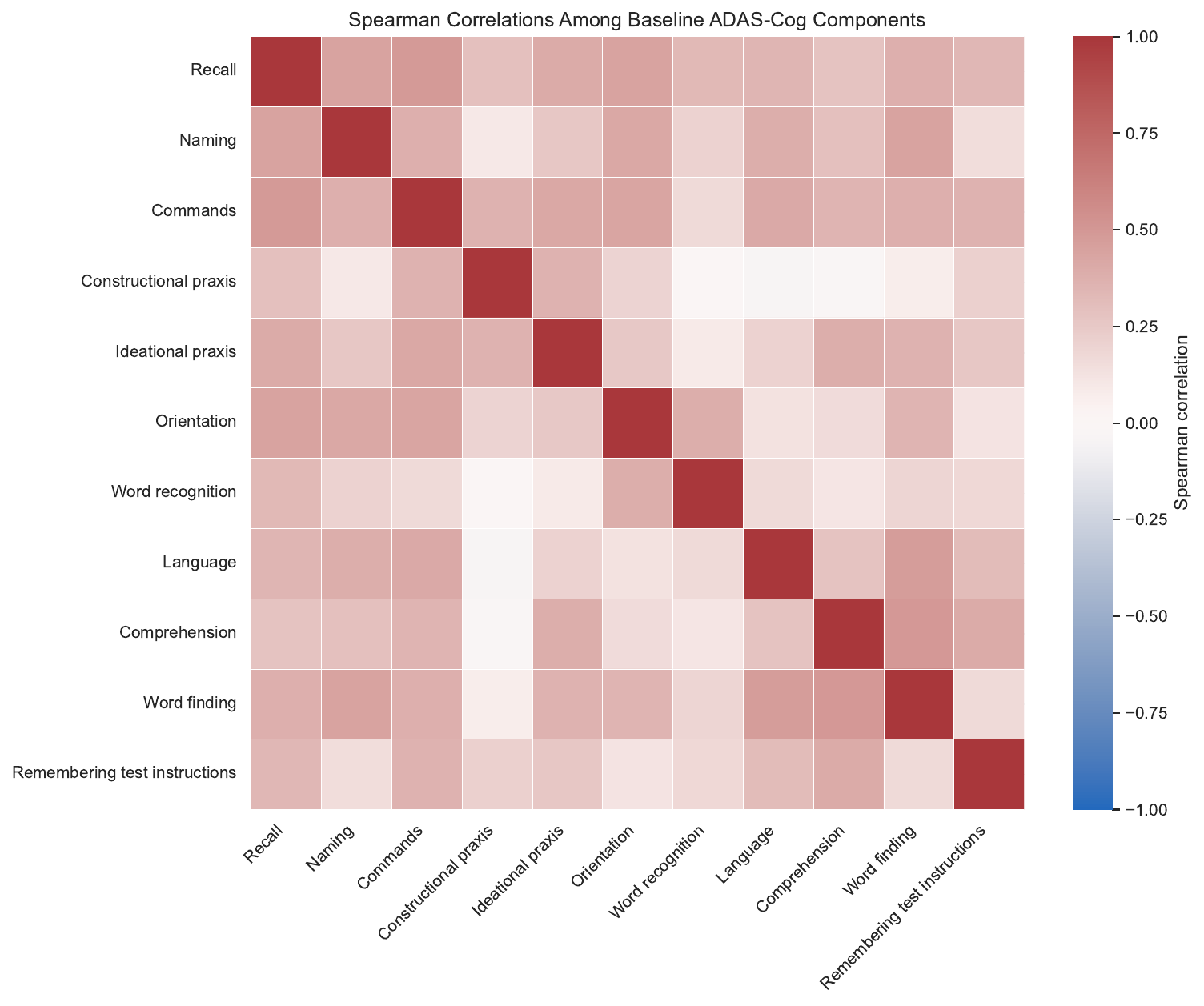}
\caption{Spearman correlations among the 11 baseline ADAS-Cog component
scores in the primary analysis cohort.}
\label{fig:baseline-domain-correlations}
\end{figure}

The 11 baseline ADAS-Cog component scores showed varying degrees of
correlation, indicating overlap among baseline cognitive measures without
complete redundancy (Figure~\ref{fig:baseline-domain-correlations}).
These correlations were descriptive and were not used for predictor
selection.

\begin{figure}[htbp]
\centering
\includegraphics[width=0.65\textwidth,height=0.60\textheight,keepaspectratio]{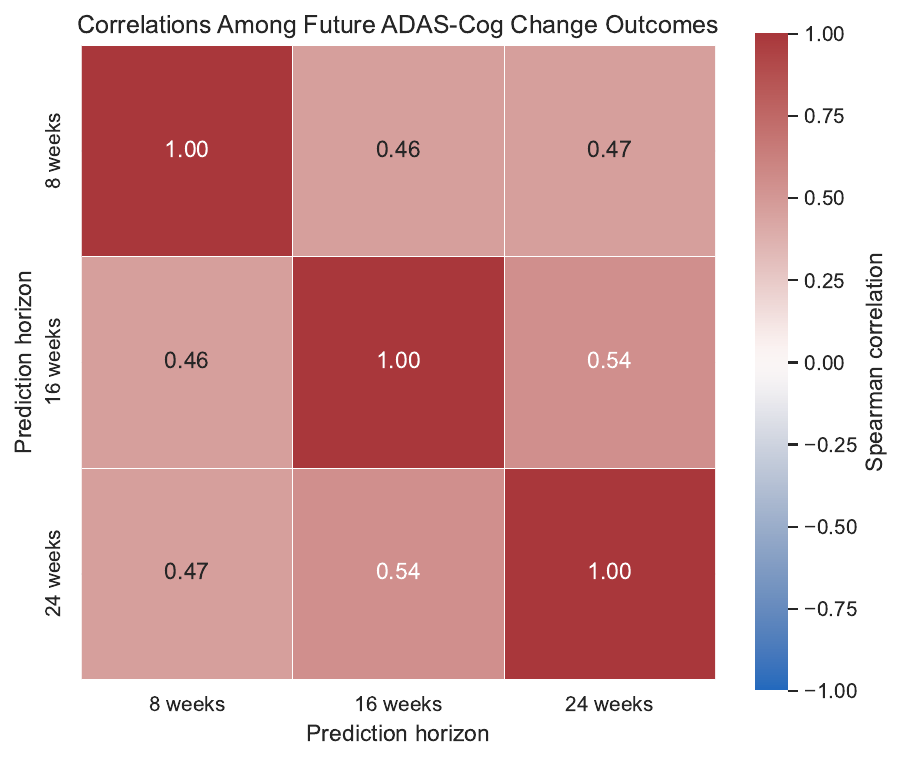}
\caption{Spearman correlations among future ADAS-Cog change outcomes
at the three prediction horizons.}
\label{fig:future-outcome-correlations}
\end{figure}

Future ADAS-Cog changes were moderately positively correlated across
prediction horizons (Figure~\ref{fig:future-outcome-correlations}).
Spearman correlation was 0.463 between the 8- and 16-week outcomes,
0.468 between the 8- and 24-week outcomes, and 0.542 between the
16- and 24-week outcomes. Thus, the three future outcomes shared
information but were not interchangeable.

\subsection{Predictor-set comparison}

The ridge predictor-set comparison showed little out-of-sample predictive
value from baseline information alone. The baseline-only model had an
overall MAE of 3.951 (SD 0.027), RMSE of 5.022 (SD 0.035), and
$R^2=-0.027$ (SD 0.014). Adding treatment group and site produced
essentially no change, with the baseline-plus-trial model yielding an
overall MAE of 3.951 (SD 0.028), RMSE of 5.022 (SD 0.034), and
$R^2=-0.027$ (SD 0.014).

Adding early total-score response produced a substantial improvement in
prediction. The early total-response ridge achieved an overall MAE of
3.067 (SD 0.037), RMSE of 3.885 (SD 0.050), and $R^2=0.384$
(SD 0.017), representing the best numerical performance among the four
ridge predictor specifications. Replacing the Week-3 and Week-5 total-score changes with the corresponding
22 component-specific changes did not improve performance. The early
domain-response ridge had an overall MAE of 3.363 (SD 0.089), RMSE of
4.269 (SD 0.145), and $R^2=0.257$ (SD 0.052). Thus, relative to the
early total-response model, the domain-response representation increased
mean MAE by approximately 0.297 points and reduced overall $R^2$ by
approximately 0.126.

\begin{table}[htbp]
\centering
\caption{Overall repeated cross-validated performance of the ridge
predictor-set specifications. Values are mean (SD) across the five
outer cross-validation repetitions.}
\label{tab:predictor-comparison}
\small
\begin{tabular}{llccc}
\toprule
Model & Predictor specification & MAE & RMSE & $R^2$ \\
\midrule
A & Baseline only
& 3.951 (0.027)
& 5.022 (0.035)
& $-0.027$ (0.014) \\

B & Baseline + trial
& 3.951 (0.028)
& 5.022 (0.034)
& $-0.027$ (0.014) \\

C & Early total response
& 3.067 (0.037)
& 3.885 (0.050)
& 0.384 (0.017) \\

D & Early domain response
& 3.363 (0.089)
& 4.269 (0.145)
& 0.257 (0.052) \\
\bottomrule
\end{tabular}
\end{table}

The same numerical ordering between the early total-response and
early domain-response specifications was observed at each prediction
horizon. At 8 weeks, the early total-response model achieved an MAE
of 2.633 compared with 3.178 for the domain-response model. At
16 weeks the corresponding MAEs were 3.285 and 3.389, and at
24 weeks they were 3.282 and 3.523
(Table~\ref{tab:total-v-domain}).

\begin{table}[htbp]
\centering
\caption{Horizon-specific repeated cross-validated performance of the
early total-response and early domain-response ridge specifications.
Values are mean (SD) across the five outer cross-validation repetitions.}
\label{tab:total-v-domain}
\small
\begin{tabular}{llccc}
\toprule
Horizon & Model & MAE & RMSE & $R^2$ \\
\midrule
8 weeks
& Early total response
& 2.633 (0.016)
& 3.523 (0.013)
& 0.530 (0.004) \\
& Early domain response
& 3.178 (0.074)
& 4.136 (0.135)
& 0.352 (0.043) \\
\midrule
16 weeks
& Early total response
& 3.285 (0.099)
& 4.053 (0.150)
& 0.227 (0.058) \\
& Early domain response
& 3.389 (0.054)
& 4.180 (0.047)
& 0.178 (0.019) \\
\midrule
24 weeks
& Early total response
& 3.282 (0.046)
& 4.080 (0.043)
& 0.363 (0.013) \\
& Early domain response
& 3.523 (0.211)
& 4.492 (0.271)
& 0.226 (0.094) \\
\bottomrule
\end{tabular}
\end{table}

\begin{figure}[htbp]
\centering
\includegraphics[width=\textwidth,height=0.60\textheight,keepaspectratio]{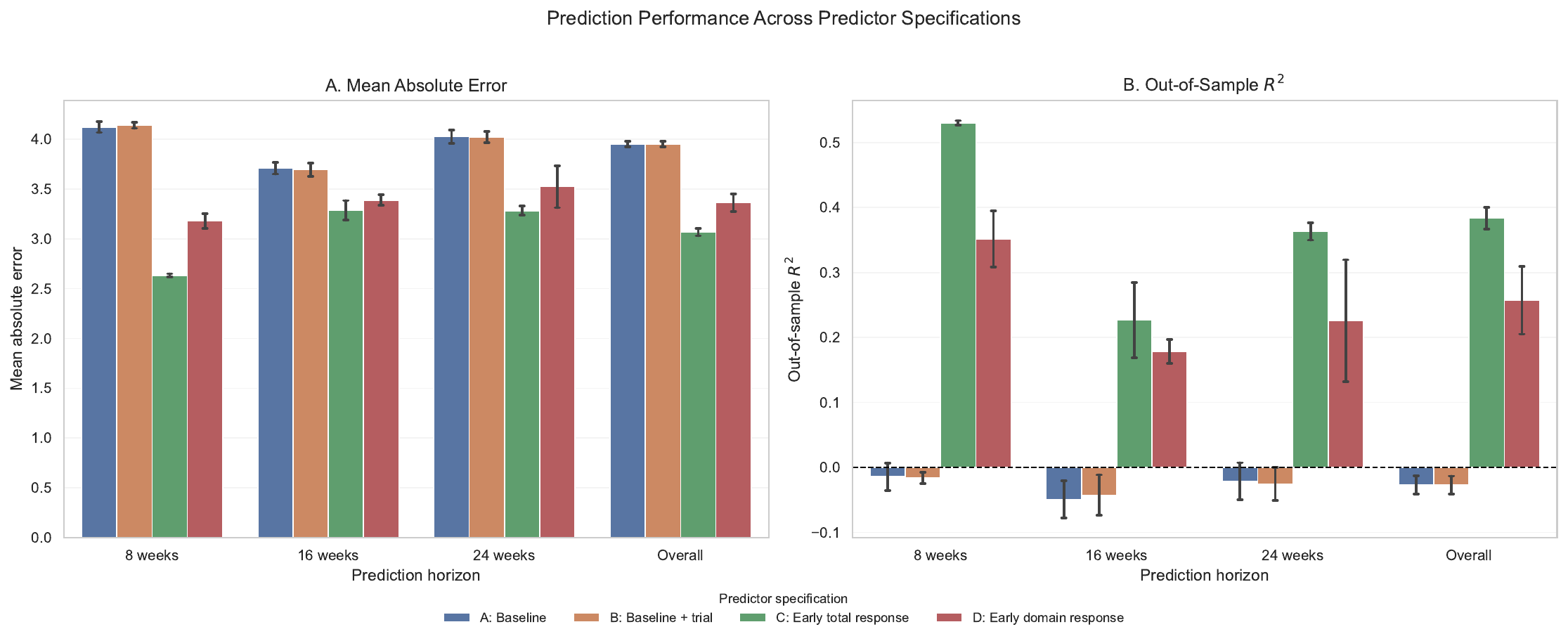}
\caption{Repeated cross-validated prediction performance across the four
ridge predictor specifications. Error bars represent the standard
deviation across the five outer cross-validation repetitions.}
\label{fig:predictor-set-performance}
\end{figure}

The predictor-set comparison is summarized in
Figure~\ref{fig:predictor-set-performance}. Overall, the results indicate
that early cognitive response provided useful predictive information,
but representing that response with individual ADAS-Cog components did
not improve prediction beyond the simpler total-score representation in
this cohort.

\subsection{Model-class comparison}

A separate analysis compared the full-domain ridge model, multi-task
elastic net, restricted random forest, and longitudinal mixed-effects
model. Among these four approaches, the mixed-effects model had the
lowest numerical overall prediction error, with MAE 3.162 (SD 0.072),
RMSE 4.091 (SD 0.105), and $R^2=0.318$ (SD 0.034).
Multi-task elastic net ranked second by overall MAE, followed by the
full-domain ridge model and restricted random forest.

The mixed-effects model also had the lowest MAE among these four
model-class comparators at each horizon: 2.920 at 8 weeks, 3.195 at
16 weeks, and 3.370 at 24 weeks. Ridge and elastic net showed broadly
similar performance, whereas the restricted random forest generally had
higher prediction error.

\begin{table}[htbp]
\centering
\caption{Repeated cross-validated performance of the four model-class
comparators. Values are mean (SD) across the five outer cross-validation
repetitions. The SDs describe variability across repeated train/test
partitions and are not confidence intervals.}
\label{tab:model-performance}
\small
\begin{tabular}{llccc}
\toprule
Model & Horizon & MAE & RMSE & $R^2$ \\
\midrule
Full-domain ridge
& 8 weeks
& 3.086 (0.056)
& 3.997 (0.049)
& 0.395 (0.015) \\
& 16 weeks
& 3.465 (0.116)
& 4.299 (0.155)
& 0.130 (0.064) \\
& 24 weeks
& 3.484 (0.101)
& 4.383 (0.137)
& 0.264 (0.045) \\
& Overall
& 3.345 (0.065)
& 4.226 (0.091)
& 0.272 (0.032) \\
\midrule
Multi-task elastic net
& 8 weeks
& 3.100 (0.076)
& 4.105 (0.061)
& 0.362 (0.019) \\
& 16 weeks
& 3.403 (0.065)
& 4.159 (0.076)
& 0.186 (0.030) \\
& 24 weeks
& 3.418 (0.100)
& 4.365 (0.136)
& 0.270 (0.045) \\
& Overall
& 3.307 (0.028)
& 4.210 (0.055)
& 0.279 (0.019) \\
\midrule
Restricted random forest
& 8 weeks
& 3.440 (0.070)
& 4.436 (0.048)
& 0.255 (0.016) \\
& 16 weeks
& 3.472 (0.059)
& 4.291 (0.089)
& 0.134 (0.036) \\
& 24 weeks
& 3.577 (0.127)
& 4.571 (0.123)
& 0.200 (0.043) \\
& Overall
& 3.496 (0.047)
& 4.433 (0.050)
& 0.201 (0.018) \\
\midrule
Mixed-effects model
& 8 weeks
& 2.920 (0.104)
& 3.896 (0.158)
& 0.425 (0.046) \\
& 16 weeks
& 3.195 (0.087)
& 4.062 (0.107)
& 0.224 (0.041) \\
& 24 weeks
& 3.370 (0.069)
& 4.316 (0.085)
& 0.287 (0.028) \\
& Overall
& 3.162 (0.072)
& 4.091 (0.105)
& 0.318 (0.034) \\
\bottomrule
\end{tabular}
\end{table}

\begin{figure}[htbp]
\centering
\includegraphics[width=\textwidth,height=0.60\textheight,keepaspectratio]{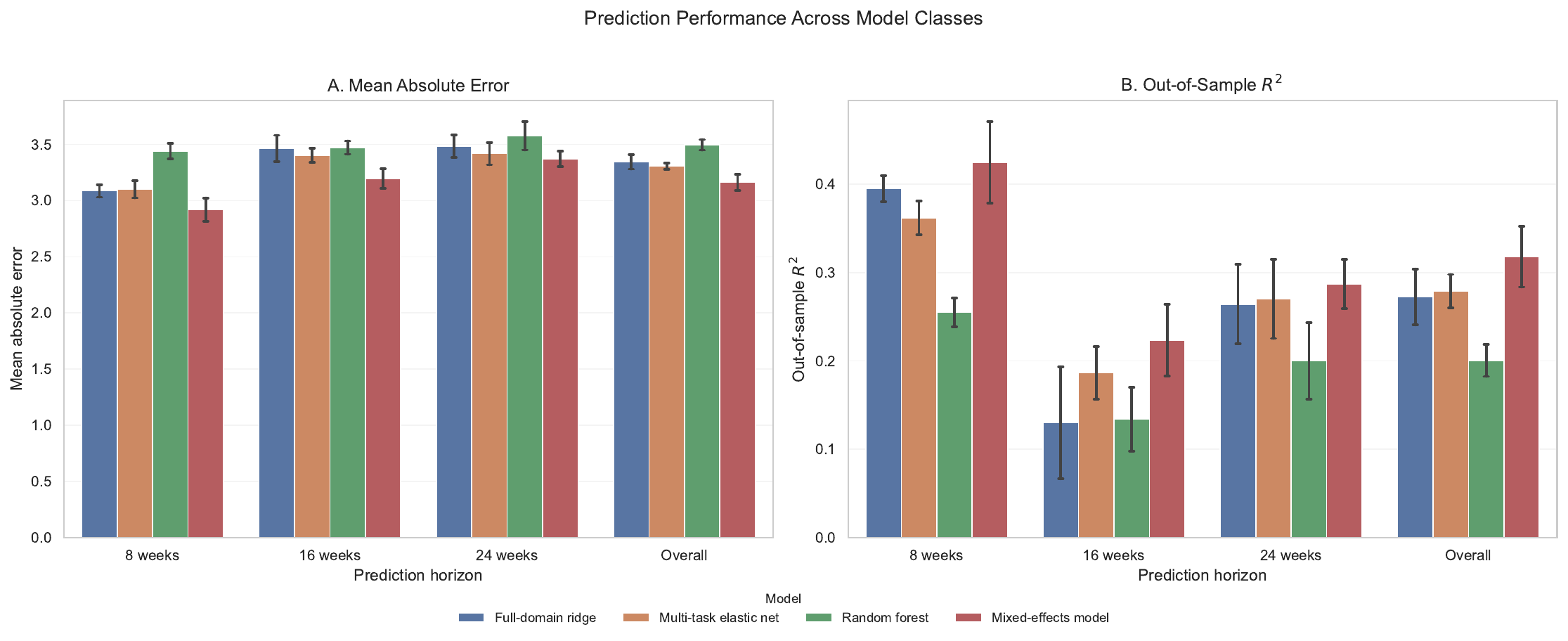}
\caption{Repeated cross-validated prediction performance across the
full-domain ridge, multi-task elastic-net, restricted random-forest,
and longitudinal mixed-effects models.}
\label{fig:model-class-performance}
\end{figure}

The model-class comparison is shown in
Figure~\ref{fig:model-class-performance}.

When the predictor-set and model-class analyses were considered together,
the early total-response ridge had the lowest numerical overall MAE
(3.067) and the highest overall $R^2$ (0.384). The mixed-effects model
was close in overall MAE at 3.162 and had an overall $R^2$ of 0.318.
At 8 weeks the early total-response ridge had lower MAE than mixed
effects (2.633 versus 2.920), whereas at 16 weeks mixed effects had
slightly lower MAE (3.195 versus 3.285). At 24 weeks the early
total-response ridge again had slightly lower MAE
(3.282 versus 3.370).

\subsection{Paired bootstrap comparisons}

Participant-level paired bootstrap analysis provided stronger evidence
for the early total-response representation than for the domain-specific
representation. Overall, the MAE difference between the early
total-response and early domain-response models was $-0.297$
(95\% CI $-0.485$ to $-0.115$), where negative values favor the
early total-response model. The bootstrap probability that the early
total-response model had lower MAE was 0.9995.

At 8 weeks, the corresponding difference was $-0.545$
(95\% CI $-0.807$ to $-0.291$). At 16 weeks the difference was
$-0.104$ (95\% CI $-0.374$ to 0.161), and at 24 weeks it was
$-0.241$ (95\% CI $-0.575$ to 0.093). Thus, the bootstrap interval
excluded zero overall and at the 8-week horizon, whereas the
16- and 24-week horizon-specific intervals remained compatible with
little or no difference.

The early total-response ridge and mixed-effects model showed much
smaller differences. Overall, the MAE difference was $-0.095$
(95\% CI $-0.221$ to 0.025), with a bootstrap probability of 0.939
that the early total-response ridge had lower MAE. At 8 weeks,
the difference favored the early total-response ridge
($-0.287$, 95\% CI $-0.535$ to $-0.042$). At 16 weeks the
difference favored mixed effects numerically
(0.090, 95\% CI $-0.121$ to 0.307), while at 24 weeks the
difference again favored the early total-response ridge numerically
($-0.089$, 95\% CI $-0.342$ to 0.156). Except for the 8-week
comparison, the intervals comparing these two approaches included zero.

\begin{table}[t]
\centering
\caption{Participant-level paired bootstrap comparison of the early
total-response ridge with the early domain-response ridge and
mixed-effects model. Differences are early total-response MAE minus
comparator MAE; negative values favor the early total-response model.}
\label{tab:primary-bootstrap}
\small
\begin{tabular}{llrrr}
\toprule
Comparator & Horizon & MAE difference & 95\% CI & $P(\Delta<0)$ \\
\midrule
Early domain response
& 8 weeks  & $-0.545$ & [$-0.807$, $-0.291$] & 1.000 \\
& 16 weeks & $-0.104$ & [$-0.374$, 0.161] & 0.775 \\
& 24 weeks & $-0.241$ & [$-0.575$, 0.093] & 0.916 \\
& Overall  & $-0.297$ & [$-0.485$, $-0.115$] & 0.9995 \\
\midrule
Mixed-effects model
& 8 weeks  & $-0.287$ & [$-0.535$, $-0.042$] & 0.988 \\
& 16 weeks & 0.090 & [$-0.121$, 0.307] & 0.206 \\
& 24 weeks & $-0.089$ & [$-0.342$, 0.156] & 0.757 \\
& Overall  & $-0.095$ & [$-0.221$, 0.025] & 0.939 \\
\bottomrule
\end{tabular}
\end{table}

\begin{figure}[htbp]
\centering
\includegraphics[width=0.90\textwidth,height=0.60\textheight,keepaspectratio]{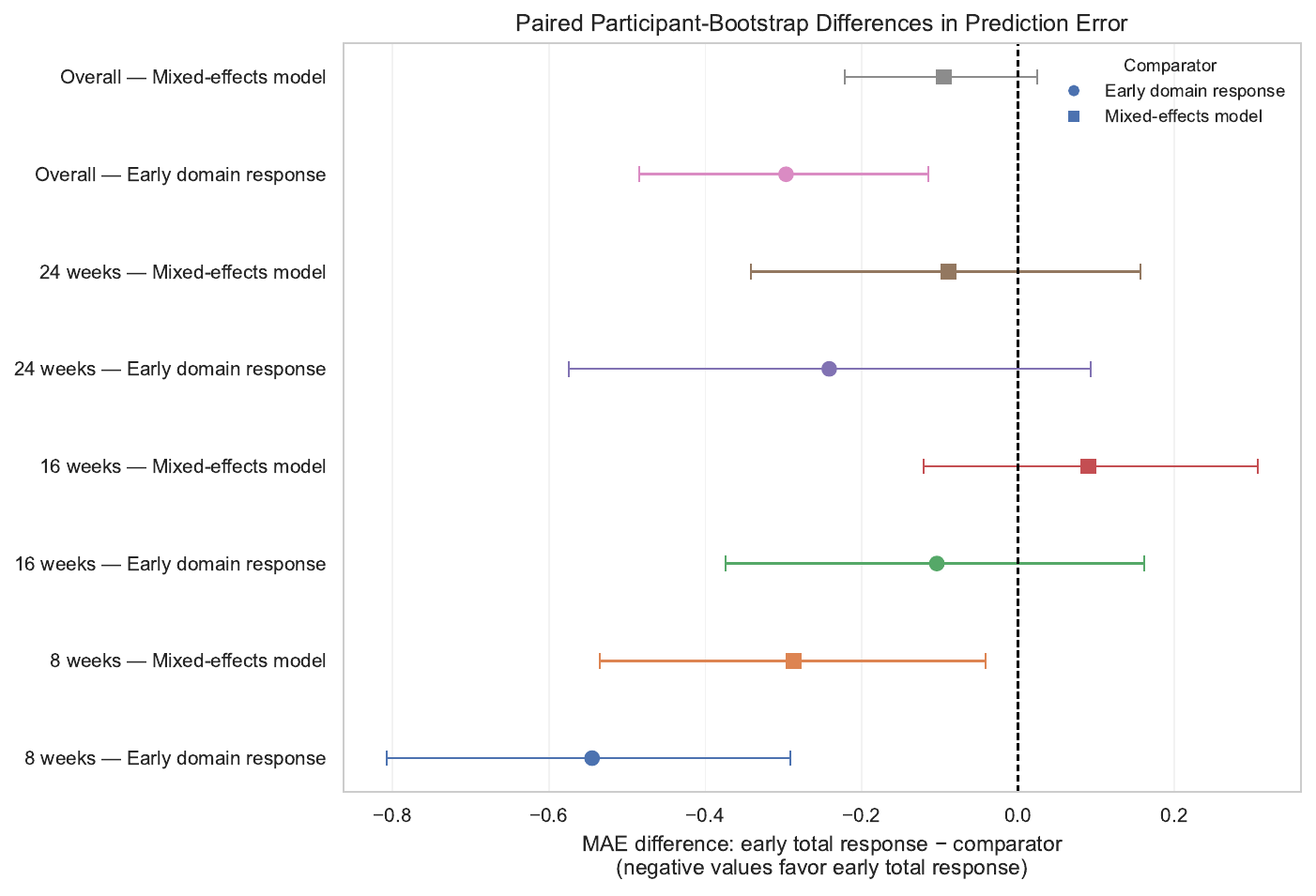}
\caption{Participant-level paired bootstrap differences in MAE for the
early total-response ridge relative to the early domain-response and
mixed-effects models. Negative values favor the early total-response
ridge. Points indicate observed MAE differences and horizontal lines
show percentile 95\% confidence intervals.}
\label{fig:bootstrap-mae}
\end{figure}

These comparisons are displayed in
Figure~\ref{fig:bootstrap-mae}.

In the secondary domain-rich model-class comparison, uncertainty
analysis also showed that numerical rankings did not always imply clear
differences. The overall full-domain ridge versus elastic-net MAE
difference was 0.038 (95\% CI $-0.057$ to 0.130), the ridge versus
random-forest difference was $-0.151$ (95\% CI $-0.323$ to 0.012),
and the ridge versus mixed-effects difference was 0.183
(95\% CI $-0.042$ to 0.410). All three overall intervals included
zero. At the 8-week horizon, the full-domain ridge had lower MAE than
the restricted random forest, with an interval excluding zero
($\Delta_{\mathrm{MAE}}=-0.354$, 95\% CI $-0.602$ to $-0.111$).

\subsection{Calibration of the early total-response ridge}

Calibration of the early total-response ridge was examined using the
mean out-of-fold prediction for each participant across the five
cross-validation repetitions. Calibration intercepts were close to zero
at all three horizons, and their 95\% confidence intervals contained
zero (Table~\ref{tab:calibration}). The calibration slope was 1.020 (95\% CI 0.819 to 1.220) at 8 weeks,
0.866 (95\% CI 0.549 to 1.184) at 16 weeks, and 1.013
(95\% CI 0.734 to 1.291) at 24 weeks. All three slope intervals
contained the ideal value of one. Therefore, within the uncertainty
of this sample, there was no clear evidence of substantial systematic
miscalibration, although the confidence intervals were relatively wide.

\begin{table}[htbp]
\centering
\caption{Calibration of the early total-response ridge using mean
out-of-fold predictions across the five repeated cross-validation runs.}
\label{tab:calibration}
\begin{tabular}{lcc}
\toprule
Horizon & Calibration intercept (95\% CI)
& Calibration slope (95\% CI) \\
\midrule
8 weeks
& 0.082 ($-0.776$, 0.939)
& 1.020 (0.819, 1.220) \\

16 weeks
& 0.109 ($-0.741$, 0.960)
& 0.866 (0.549, 1.184) \\

24 weeks
& 0.004 ($-0.953$, 0.962)
& 1.013 (0.734, 1.291) \\
\bottomrule
\end{tabular}
\end{table}

\begin{figure}[htbp]
\centering
\includegraphics[width=\textwidth,height=0.60\textheight,keepaspectratio]{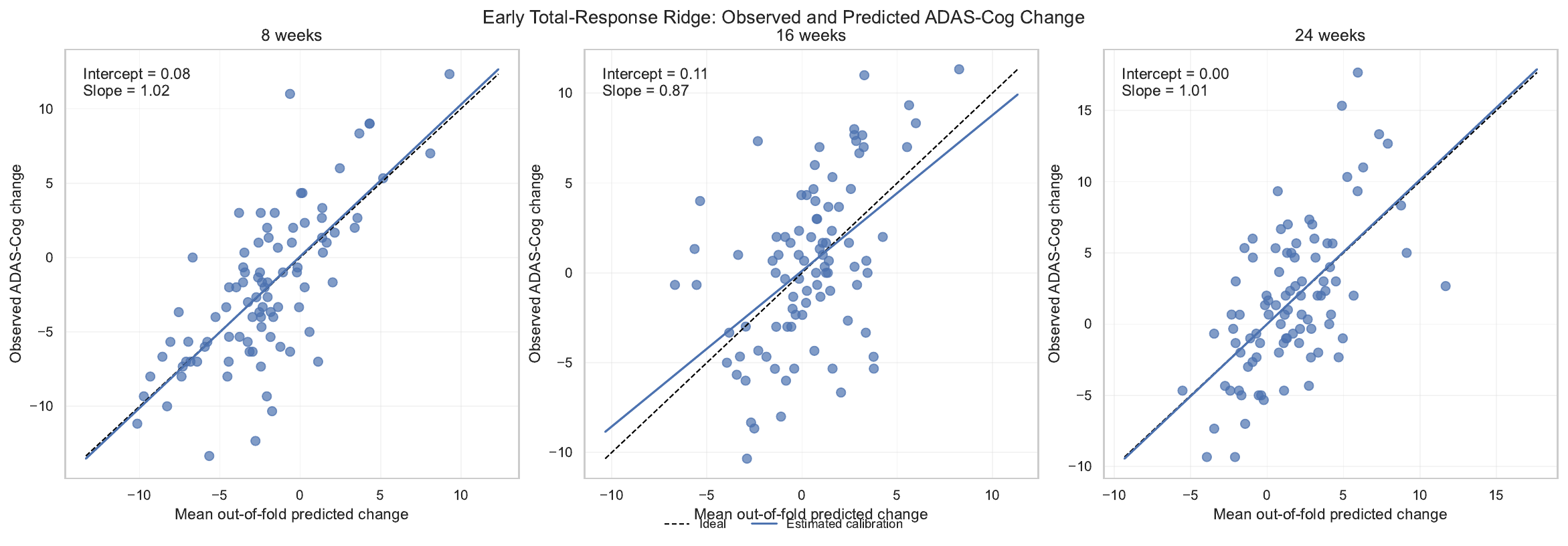}
\caption{Observed versus mean out-of-fold predicted ADAS-Cog change for
the early total-response ridge at the three prediction horizons.
Dashed lines indicate ideal calibration and solid lines indicate the
estimated calibration relationship.}
\label{fig:calibration}
\end{figure}

Observed-versus-predicted values and estimated calibration lines are
shown in Figure~\ref{fig:calibration}.

\subsection{Sensitivity analyses}

Two sensitivity analyses were conducted for the primary early
total-response ridge model. First, to evaluate whether predictive performance was driven by
mathematical coupling from using the same baseline ADAS-Cog measurement
in both the early change predictors and the future change outcomes,
the model was refitted using absolute follow-up ADAS-Cog total scores
as the outcomes. The predictor set was otherwise unchanged and included
age, sex, baseline ADAS-Cog total, treatment group, site, Week-3
total-score change, and Week-5 total-score change. Prediction remained substantial with the alternative outcome definition.
The model achieved an overall MAE of 3.169 (SD 0.100), RMSE of
4.003 (SD 0.130), and $R^2=0.838$ (SD 0.011). Horizon-specific
$R^2$ values were 0.847 at 8 weeks, 0.809 at 16 weeks, and
0.856 at 24 weeks. Because absolute follow-up scores and
change-from-baseline outcomes have different variance structures, and
because baseline ADAS-Cog total was explicitly included as a predictor,
these $R^2$ values are not directly comparable with those from the
primary change-score analysis. Nevertheless, preservation of predictive
performance under the alternative outcome definition reduces concern
that the primary results were attributable solely to shared-baseline
mathematical coupling.

Second, because the Week-3 assessment occurred after completion of
treatment for R2 but during the intervention period for R4 and S4,
the primary model was refitted after removing Week-3 total-score change.
The resulting specification retained age, sex, baseline ADAS-Cog total,
treatment group, site, and Week-5 total-score change. Removing Week-3 response reduced predictive performance. The primary
early total-response ridge had an overall MAE of 3.067, RMSE of 3.885,
and $R^2=0.384$, whereas the model without Week-3 response had an
overall MAE of 3.392, RMSE of 4.204, and $R^2=0.279$. The same
pattern was observed at each horizon. MAE increased from 2.633 to
2.960 at 8 weeks, from 3.285 to 3.570 at 16 weeks, and from
3.282 to 3.647 at 24 weeks. These results indicate that Week-5
response alone retained predictive information, but Week-3 total-score
response contributed additional predictive information to the primary
model. The complete-case restriction was also examined descriptively.
Participants excluded from the primary analysis were somewhat older
and had a higher mean baseline ADAS-Cog score than included participants.
Outcome availability among the 101 Week-5 landmark participants remained
high, with 99, 95, and 94 participants having observed outcomes at
8, 16, and 24 weeks post-treatment, respectively.

\FloatBarrier
\section{Discussion}

The principal finding of this study was that early cognitive response after the Week-5 landmark substantially improved prediction of later ADAS-Cog change, but representing that response at the level of individual ADAS-Cog components did not improve prediction beyond the simpler total-score representation. Baseline demographic and cognitive information alone provided little out-of-sample predictive value, with negative overall $R^2$ for both the baseline-only and baseline-plus-trial ridge models. In contrast, adding Week-3 and Week-5 total-score changes reduced overall MAE to 3.067 and increased overall $R^2$ to 0.384. Replacing these two aggregate response measures with 22 component-specific change variables worsened performance, with overall MAE increasing to 3.363 and $R^2$ decreasing to 0.257. The paired participant bootstrap supported this difference overall: the early total-response ridge had an MAE 0.297 points lower than the early domain-response ridge, with a 95\% confidence interval of $-0.485$ to $-0.115$. These results support the predictive value of early cognitive response, but they do not support the stronger hypothesis that a domain-specific representation adds predictive value beyond the aggregate ADAS-Cog response in this sample.

Modeling ADAS-Cog components separately remains clinically and psychometrically reasonable because the total score combines tasks assessing memory, orientation, language, comprehension, and praxis, and the individual components have different measurement characteristics \citep{cano2010parts}. Thus, domain-level modeling could theoretically distinguish participants with similar total scores but different patterns of cognitive response. In the present data, however, this additional resolution did not improve out-of-sample prediction. One plausible explanation is that aggregation of component scores reduces measurement noise, whereas replacing two early total-score changes with 22 component-level changes substantially increases model dimensionality relative to a sample of only 91 participants. Regularization can reduce but does not eliminate the instability associated with fitting many predictors in small samples \citep{riley2019continuous}. The present findings therefore should not be interpreted as evidence that individual ADAS-Cog domains are clinically unimportant. Rather, decomposition of the early response into individual components did not improve the specific prediction task evaluated here.

The comparison among model classes provided a complementary result. Among the domain-rich model specifications, the longitudinal mixed-effects model had the lowest numerical overall prediction error, with MAE 3.162 and $R^2=0.318$, followed by multi-task elastic net, full-domain ridge regression, and restricted random forest. However, when all evaluated candidate models were considered together, the simpler early total-response ridge had the lowest numerical overall MAE, 3.067, and the highest overall $R^2$, 0.384. The overall difference between the early total-response ridge and the mixed-effects model was small and uncertain in the paired bootstrap analysis ($\Delta\mathrm{MAE}=-0.095$, 95\% CI $-0.221$ to 0.025). At 8 weeks, the early total-response ridge had lower error, whereas the mixed-effects model was slightly better numerically at 16 weeks and the two approaches were close at 24 weeks. These results do not support a claim that one of the two approaches is definitively superior. Instead, they suggest that a relatively simple regularized early-response model and a structured longitudinal model can provide similar predictive performance using different representations of the available information.

The similarity between full-domain ridge regression and multi-task elastic net also suggests that imposing shared sparsity across the three future outcomes did not provide a substantial advantage over conventional coefficient shrinkage. The future ADAS-Cog changes were moderately correlated, with Spearman correlations of 0.463 between the 8- and 16-week outcomes, 0.468 between the 8- and 24-week outcomes, and 0.542 between the 16- and 24-week outcomes. Thus, the three horizons shared information but were not interchangeable. Multi-task elastic net was therefore a reasonable candidate because it can encourage joint predictor selection across outcomes, but its near-equivalence to ridge suggests that strong shrinkage was sufficient in this sample. This is broadly consistent with previous work showing competitive performance of regularized linear methods for predicting cognitive decline in Alzheimer disease \citep{imani2021comparison,zou2005elasticnet}. The restricted random forest did not improve prediction over the regularized linear or mixed-effects approaches. This may reflect the limited sample available for reliably estimating nonlinearities and interactions. Although tree-based ensembles are capable of modeling complex response surfaces \citep{breiman2001random}, such flexibility can be difficult to exploit when the sample size is small relative to the predictor space. The restricted tuning strategy used here was intended to limit overfitting, yet the random forest still had the largest overall prediction error among the domain-rich models. This does not demonstrate that nonlinear relationships are absent; rather, the present dataset did not provide evidence that modeling them using a random forest improved prediction.

The mixed-effects model should also be interpreted in relation to the prediction task. It used grouped memory, language, and praxis composites and explicitly modeled within-participant dependence using a participant-specific random intercept, whereas the supervised domain-rich models used all individual components. Moreover, participant-specific random effects could not be estimated for held-out participants who were absent from the training data. Predictions for those individuals therefore relied on the fixed-effects component only. This evaluation represents prediction for previously unseen participants rather than conditional prediction for individuals already contributing longitudinal outcome data. The performance of the mixed-effects model should therefore not be interpreted as evidence for or against its suitability for longitudinal inference; instead, it reflects how that prespecified longitudinal formulation performed as a forecasting model under participant-level internal validation.

Calibration provided additional information beyond average error measures. For the early total-response ridge, calibration slopes were 1.020, 0.866, and 1.013 at the 8-, 16-, and 24-week horizons, respectively, and calibration intercepts were close to zero. The 95\% confidence intervals for all three slopes included the ideal value of one. Thus, there was no clear evidence of major systematic miscalibration within the analyzed sample, although the intervals were relatively wide. Calibration is important because acceptable MAE or $R^2$ does not guarantee that predictions are appropriately scaled across the outcome range \citep{vancalster2019calibration}. These results remain internally estimated, however, and calibration may change when the model is applied to a different population \citep{ramspek2021external}.

The prediction task should be distinguished clearly from the treatment-effect question addressed by the parent randomized trial. The parent study reported improvement in both active and sham groups and did not demonstrate a clear active-versus-sham advantage \citep{moussavi2024trial}. The present analysis addressed a different question: whether information available by the Week-5 landmark could forecast subsequent cognitive change. Treatment group and early cognitive response were predictors rather than causal treatment-effect estimates. Predictive association does not establish causation, and neither the treatment variables nor the exploratory domain coefficients should be interpreted as evidence that rTMS caused the observed future cognitive changes \citep{shmueli2010explain}. Because predictors included observations through Week 5, the model is an early-response updating tool rather than a pretreatment patient-selection model. It therefore cannot determine before treatment whether a patient should receive rTMS. Instead, it updates the prediction after early cognitive response has already been observed. This structure is related to landmark prediction, in which information available at a defined follow-up point is used to forecast subsequent outcomes \citep{vanhouwelingen2007landmark}. The strong difference between the baseline-only models and the early-response model reinforces this distinction: most of the useful predictive information in the present analysis emerged after early follow-up measurements became available.

The sensitivity analyses further clarified this interpretation. First, the primary early total-response model remained predictive when absolute follow-up ADAS-Cog scores were used instead of change-from-baseline outcomes. Under this alternative outcome definition, overall MAE was 3.169, RMSE was 4.003, and $R^2$ was 0.838. The absolute-score $R^2$ should not be compared directly with the primary change-score $R^2$ because the two outcomes have different variance structures and baseline ADAS-Cog was explicitly included as a predictor. Nevertheless, preservation of predictive performance when the future outcome no longer contained the same baseline subtraction as the early change predictors reduces concern that the primary result was produced solely by shared-baseline mathematical coupling.

Second, removing Week-3 total-score change reduced predictive performance. The primary early total-response model had an overall MAE of 3.067 and $R^2=0.384$, whereas the model using baseline information and Week-5 response without Week-3 response had an overall MAE of 3.392 and $R^2=0.279$. The reduction in performance was also observed at each individual horizon. Thus, Week-5 response alone retained meaningful predictive information, but Week-3 response contributed additional predictive value. Interpretation of this result requires caution because Week 3 represented different stages of treatment across the randomized groups: R2 participants had completed their 2-week active intervention, whereas R4 and S4 participants remained within their 4-week intervention period. Consequently, the additional information carried by Week-3 response cannot be interpreted as a uniform post-treatment effect across all participants. Future studies could address this issue by defining early-response landmarks relative to treatment completion rather than calendar week.

The pattern of missing follow-up data is also relevant. Of the 112 randomized participants, 101 contributed complete information through the Week-5 landmark and 91 had complete outcomes at all three future horizons. Among the landmark cohort, outcome data were available for 99 participants at 8 weeks, 95 at 16 weeks, and 94 at 24 weeks. Recruitment and retention are recognized challenges in longitudinal Alzheimer disease studies \citep{grill2010retention}, and large forecasting datasets such as TADPOLE also contain incomplete longitudinal measurements and irregular follow-up \citep{marinescu2021tadpole}. However, Figure~\ref{fig:availability} cannot determine why individual visits were missing or whether missingness was associated with cognitive outcome. Participants excluded from the primary complete-case analysis were somewhat older and had higher baseline ADAS-Cog scores than those included. Complete-case analysis can introduce bias when missingness depends on participant characteristics or outcomes \citep{sterne2009missing}. The present models should therefore not be interpreted as methods for imputing missing assessments or predicting dropout.

The magnitude of prediction error should also be interpreted cautiously. The early total-response model had an overall MAE of 3.067 ADAS-Cog points and RMSE of 3.885 points. Changes of approximately three ADAS-Cog points have been discussed when considering clinically relevant change \citep{schrag2012clinically}, but a clinical-change threshold and a prediction-error metric represent different concepts. An MAE of approximately three points does not imply that all individual predictions are within three points of the observed outcome, nor does it establish that predictions are sufficiently accurate for clinical decision-making. Clinical usefulness would require independent validation, assessment of calibration in external populations, and evaluation of whether use of the prediction changes patient management or outcomes \citep{vancalster2019calibration,ramspek2021external}. Previous Alzheimer disease prediction studies provide useful context, although direct numerical comparisons are limited because populations, predictors, targets, horizons, and validation procedures differ. Prakash and colleagues predicted absolute future ADAS-Cog scores from baseline multimodal information over 12--36 months and reported MAEs of approximately 3.5, 3.7, and 4.6 when baseline ADAS-Cog was included \citep{prakash2021quantitative}. The TADPOLE challenge demonstrated the difficulty of prospectively forecasting ADAS-Cog13 in a broader Alzheimer disease population \citep{marinescu2021tadpole}. Fisher and colleagues studied longitudinal ADAS-Cog11 trajectories in a substantially larger placebo-arm cohort and found that a generative trajectory model performed comparably with endpoint-specific supervised approaches \citep{fisher2019forecasting}. In the rTMS setting, Lithgow and colleagues evaluated pretreatment classification of cognitive improvement, but their binary prediction target and predictor modality differ from the continuous post-landmark forecasting problem considered here \citep{lithgow2022baseline}. A concise comparison is provided in Table~\ref{tab:literature}.

\begin{table}[htbp]
\centering
\caption{Selected studies of cognitive outcome prediction in Alzheimer disease.}
\label{tab:literature}
\small
\begin{tabular}{p{0.17\textwidth}p{0.38\textwidth}p{0.35\textwidth}}
\toprule
Study & Prediction setting & Relation to the present study \\
\midrule

Present study
& rTMS trial; ADAS-Cog change at 8, 16, and 24 weeks post-treatment from a Week-5 landmark; overall MAE 3.067 and $R^2=0.384$
& Early-response updating; aggregate total-score response outperformed the domain-specific response representation \\

\addlinespace

Prakash et al. \citep{prakash2021quantitative}
& ADNI; absolute future ADAS-Cog prediction over 12--36 months
& Longer prediction horizons and different outcome definition \\

\addlinespace

Marinescu et al. \citep{marinescu2021tadpole}
& TADPOLE; prospective longitudinal ADAS-Cog13 forecasting
& Larger and more heterogeneous prospective prediction setting \\

\addlinespace

Fisher et al. \citep{fisher2019forecasting}
& 1,909 placebo-arm participants; longitudinal ADAS-Cog11 trajectories
& Much larger cohort and untreated-progression setting \\

\addlinespace

Lithgow et al. \citep{lithgow2022baseline}
& rTMS cohort; pretreatment binary classification of improvement
& Different target and predictor modality; not directly comparable \\

\bottomrule
\end{tabular}
\end{table}

Several features strengthen the analysis. Validation was performed at the participant level, preventing measurements from the same individual from appearing in both training and test data. Hyperparameter tuning was nested within the outer cross-validation procedure, reducing selection-related optimism \citep{cawley2010overfitting}. The same outer splits were used across competing models, permitting paired comparisons. Repetition of the outer validation reduced dependence on one arbitrary partition, while participant-level bootstrap comparisons quantified uncertainty in model differences. The analysis also evaluated calibration, directly compared aggregate and component-level early response, examined the impact of Week-3 information, and addressed shared-baseline mathematical coupling using an alternative absolute-outcome formulation.

Several limitations constrain interpretation. First, the primary complete-case cohort contained only 91 participants. The domain-rich models contained many predictors relative to this sample size, and regularization cannot eliminate all uncertainty associated with small-sample model development \citep{riley2019continuous}. Small samples can also produce substantial variability in cross-validated performance estimates \citep{varoquaux2018crossvalidation}. Second, the standard deviations across repeated cross-validation runs are descriptive measures of split-to-split variability rather than independent standard errors because observations and training sets overlap across folds and repetitions \citep{bengio2004variance}. The paired bootstrap provides a complementary assessment of uncertainty, but the models were not refitted within every bootstrap replicate.

Third, complete-case analysis excluded 21 of the 112 randomized participants, and the excluded participants differed descriptively in age and baseline cognitive score. Missingness may therefore have affected the composition of the analyzed cohort \citep{sterne2009missing}. Fourth, all model development and validation were conducted within one supplied two-site trial dataset. Internal validation estimates reproducibility within the development population but cannot establish transportability to independent populations \citep{ramspek2021external}. Differences in patient characteristics, stimulation protocols, measurement procedures, or predictor--outcome relationships may reduce performance when a model is transported to another setting \citep{subbaswamy2020shift}. External validation is therefore required before clinical application.

Fifth, the early total-response ridge and mixed-effects model did not use identical predictor representations. Their comparison is useful for evaluating plausible forecasting strategies but should not be interpreted as a pure algorithm comparison with identical input information. Sixth, the domain-level coefficient analyses involve correlated and regularized predictors and therefore represent conditional predictive associations rather than independent biological effects. Finally, the present study evaluates prediction rather than treatment efficacy or clinical utility. 

Taken together, the findings shift the interpretation of this study from domain-specific prediction toward early-response updating. Early change in the overall ADAS-Cog total contained substantial information about subsequent cognitive change, whereas decomposing the same early response into individual ADAS-Cog components did not improve prediction in this sample. Week-3 information contributed additional predictive value beyond Week-5 response alone, and predictive performance remained substantial when absolute future ADAS-Cog scores were used instead of change-from-baseline outcomes. The longitudinal mixed-effects model provided competitive performance, but its overall difference from the early total-response ridge remained uncertain. These findings support a parsimonious early-response forecasting strategy while emphasizing the need for larger samples, independent external validation, and prospective evaluation before clinical use.
\FloatBarrier
\section{Conclusion}

This study evaluated whether early cognitive response after rTMS could improve prediction of later ADAS-Cog change in participants with Alzheimer disease. The main finding was that early total-score response provided substantially more predictive information than baseline demographic, cognitive, treatment, and site variables alone. The early total-response ridge model achieved the best numerical overall performance among the evaluated candidate models, with an overall MAE of 3.067 and $R^2=0.384$. Decomposing early cognitive response into individual ADAS-Cog components did not improve prediction. The domain-response ridge produced higher prediction error and lower explained variance than the simpler total-response model, and the paired bootstrap supported an overall advantage for the aggregate early-response representation. These findings suggest that, in this sample, the predictive information contained in early cognitive change was captured more effectively by the overall ADAS-Cog response than by a higher-dimensional representation of individual component changes. The longitudinal mixed-effects model provided competitive performance, with only a small and uncertain overall difference from the early total-response ridge. This indicates that both a parsimonious regularized model and a structured longitudinal approach can provide useful forecasts of later cognitive change, although they represent the available information differently.

Sensitivity analyses supported the overall predictive findings. Prediction remained substantial when absolute future ADAS-Cog scores were used instead of change-from-baseline outcomes, reducing concern that the primary results were attributable solely to mathematical coupling through the shared baseline measurement. Removing Week-3 total-score response reduced performance, indicating that Week-3 information contributed additional predictive value beyond Week-5 response alone. Because Week 3 represented different treatment stages across the randomized groups, however, this contribution should be interpreted as predictive rather than as evidence of a uniform treatment effect. The present model should be viewed as an early-response updating tool rather than a pretreatment treatment-selection model. Its predictions use information available through the Week-5 landmark and therefore cannot determine before treatment which patients are likely to benefit from rTMS. Furthermore, the analysis was based on a small complete-case cohort and relied on internal validation within a single supplied trial dataset. The results do not establish treatment efficacy, external transportability, or clinical utility.

Overall, the findings support a relatively simple early-response forecasting strategy in which aggregate ADAS-Cog change observed during the early treatment period is used to update predictions of subsequent cognitive trajectory. Larger independent cohorts, external validation, and prospective evaluation will be required before such a model could be considered for clinical application.

\section*{Acknowledgements}
The authors gratefully acknowledge the institutional and research support provided by Prof. Hina Shaheen, which significantly contributed to the successful completion of this study.

\section*{Data and Code Availability}

The analysis code supporting the findings of this study
is available in the GitHub repository
\href{https://github.com/Alamgir9/rTMS-ADAS-Cog-Prediction}
{rTMS-ADAS-Cog-Prediction}.
The participant-level dataset is not included in the public repository.

\section*{Funding}
The research did not receive any specific grant from funding agencies in the public, commercial or not-for-profit sectors.

\FloatBarrier

%
%
%
%

\bibliographystyle{plainnat}
\bibliography{references_rtms}

\end{document}